\documentclass[final,5p,times,twocolumn,numbers,sort, compress]{elsarticle}
\usepackage{amssymb}
\usepackage{lipsum}
\usepackage{xcolor}
\usepackage{mathtools}
\usepackage{amsmath}
\usepackage[utf8]{inputenc}
\usepackage{subfig}
\usepackage{float}

\begin{document}

\begin{frontmatter}

\title{Rare events for low energy domain in bouncing ball model}

\author{$^1$Edson D. Leonel,  $^2$Diego F. M. Oliveira}
\address{{$^1$Departamento de F\'isica, UNESP - Universidade Estadual Paulista, Av. 24A, 1515, Bela Vista, Rio Claro, 13506-900, Sao Paulo, Brazil\\
$^2$School of Electrical Engineering and Computer Science - University of North Dakota, Grand Forks, Avenue Stop 8357, 58202, North Dakota, USA.}}

\begin{abstract}

The probability distribution for multiple collisions observed in the chaotic low energy domain in the bouncing ball model is shown to be scaling invariant concerning the control parameters. The model considers the dynamics of a bouncing ball particle colliding elastically with two rigid walls. One is fixed, and the other one moves periodically in time. The dynamics is described by a two-dimensional mapping for the variables velocity of the particle and phase of the moving wall. For a specific combination of velocity and phase, the particle may experience a type of rare collision named successive collisions. We show that a power law describes the probability distribution of the multiple impacts and is scaling invariant to the control parameter.
\end{abstract}


\end{frontmatter}




\section{Introduction}

Nonlinear systems are common in many scientific areas, including natural phenomena and social processes. These systems are known for their sensitivity to initial conditions, which leads to complex and unpredictable behaviors. One interesting aspect of nonlinear systems is the occurrence of rare events—uncommon but important occurrences that can greatly influence the overall behavior of the system.

Although rare events have a low probability, they can have significant and important consequences. These events arise from complex internal interactions, making them difficult to predict. Unlike linear systems, which are more predictable, nonlinear systems can show a wide range of behaviors, such as chaos and sudden changes in state. Small perturbations in these systems can lead to large changes, often triggered by random fluctuations or deterministic instabilities. Rare events  are linked to phenomena like tipping points \cite{lenton2013environmental}, extreme weather events \cite{ummenhofer2017extreme}, critical transitions \cite{bianconi2018rare,perez2019sampling}, and catastrophic failures \cite{komljenovic2016risks}. For example, in climate science, understanding rare events like heatwaves or floods is essential for predicting the impacts of climate change \cite{national2016attribution,katz2002statistics}, in ecological systems, a rare event might be the sudden collapse of a species population \cite{keeling2008modeling};  epidemiology, rare events such as disease outbreaks are crucial for public health strategies \cite{keeling2008modeling}; in financial markets, it could be a sudden crash \cite{taleb2010black}; in engineering, understanding rare events like structural collapses is vital for improving safety \cite{ditlevsen2010tipping};   in social networks, it might involve the rapid spread of misinformation \cite{bessi2017statistical,lu2019clustering}. These events are closely related to the nonlinear nature of the system, where small changes can have large effects. In stochastic nonlinear systems, rare events are often studied through concepts like exit times or first-passage times, which measure the time it takes for the system to reach a particular state. These rare events are not only important for understanding the full behavior of nonlinear dynamical systems but also for developing strategies to mitigate their potentially devastating effects. Analytical methods like extreme value theory, stochastic modeling, and numerical simulations are often used to study these rare events, helping to understand their mechanisms and anticipate their effects. 
 The probability of such events depends on transition rates, which are calculated using methods such as stochastic differential equations, Monte Carlo simulations, and large deviation theory. Understanding rare events is essential for improving the resilience and stability of both natural and engineered systems.

In this paper, we will revisit a problem that originated in 1949 when Enrico Fermi proposed a theory to describe high-energy cosmic rays. Fermi suggested that a charged particle could be accelerated by interacting with time-dependent magnetic structures \cite{fermi1949origin}. After Fermi's groundbreaking paper, various versions of this theoretical model have been developed, incorporating factors like external fields \cite{liang2024dynamics}, dissipation \cite{luna1990regular,diego2011parameter}, quantum \cite{jose1986study,dembinski1993quantum} and relativistic effects \cite{pustyl1995poincare}. 

One of the most studied versions of this model is the well-known Fermi–Ulam model (FUM) \cite{leonel2004dynamical}. This model describes a classical particle confined between two rigid walls, where one wall is fixed, and the other oscillates periodically in time. In the non-dissipative version, all collisions with the walls are assumed to be elastic. The phase space of the FUM has a mixed structure, meaning that depending on the combination of control parameters and initial conditions, one can observe invariant spanning curves, chaotic seas, and periodic islands. The invariant spanning curves prevent the particle from gaining unlimited energy in the chaotic sea, thus preventing Fermi acceleration. Critical exponents and scaling invariance in the FUM have also been characterized. Rare events in the FUM occur when a particle, due to specific initial conditions and control parameters, enters the collision zone and experiences multiple collisions with the moving wall. These rare and significant energy-transfer events can lead to large changes in the particle's dynamics. Understanding these rare collisions and their effects is important for a complete analysis of particle behavior in the FUM, as they highlight the complex relationship between deterministic chaos and stochastic fluctuations \cite{lichtenberg2013regular}.

This paper is organized as follows: Section \ref{sec2} describes the one-dimensional Fermi-Ulam model. In Section \ref{sec3} we present our numerical results. Finally, Section \ref{conc} presents the discussions and conclusions.

\section{The model and some dynamical properties}
\label{sec2}

The model comprises a classical particle of mass $m$ confined to moving between two rigid walls. One of them is fixed in $x=l$, and the one is described by  $x_w(t)=\varepsilon\cos(\omega t)$ where  $\varepsilon$ is the amplitude of the oscillating wall, $\omega$ is the frequency of oscillation and $t$ is the time. 

The dynamics can be described by a two-dimensional mapping for the velocity of the particle and the instant of the impact $(v,t)$ from the collision $n$ to the impact $(n+1)$ measured at the moving wall. Two different types of collisions are observed, namely: (i) the direct with the particle hitting one wall followed by the other one or (ii) the indirect (multiple) collisions when the particle has consequent impacts with the moving wall inside of a collision zone $x\in[-\varepsilon,\varepsilon]$.

A straightforward construction of the mapping can be seen in Refs. \cite{lichtenberg2013regular,leonel2021scaling} and we leave only the mapping in the dimensionless variables here. Therefore, we use $V=v/(\omega l)$, $\epsilon=\varepsilon/l$ and measure the time in terms of the number of oscillations of the moving wall, i.e., measuring phases $\phi=\omega t$. For these variables, the mapping is written as
\begin{equation}
T:\left\{\begin{array}{ll}
\phi_{n+1}=[\phi_n+\Delta T_n]~{\rm mod}~2\pi,\\
V_{n+1}=V^*-2\epsilon\sin(\phi_{n+1}),\\
\end{array}
\right.
\label{fum_eq}
\end{equation}
where $V^*$ and $\Delta T_n$ are determined according to the type of collision. For the successive collisions, $V^*=-V_n$ and $\Delta T_n=\phi_c$ where $\phi_c$ is obtained from the solution of $G(\phi_c)=0$ with
\begin{equation}
G(\phi_c)=\epsilon\cos(\phi_n+\phi_c)-\epsilon\cos(\phi_n)-V_n\phi_c,
\label{c10_eq10}
\end{equation}
and $\phi_c\in(0,2\pi]$. The case of indirect collisions gives $V^*=V_n$ and $\Delta T_n=\phi_r+\phi_l+\phi_c$ with $\phi_c$ obtained from the solution of
\begin{equation}
F(\phi_c)=\epsilon\cos(\phi_n+\phi_r+\phi_l+\phi_c)-\epsilon+V_n\phi_c=0,
\label{c10_eq11}
\end{equation}
where
\begin{eqnarray}
\phi_r={{1-\epsilon\cos(\phi_n)}\over{V_n}},~~~~  \phi_l={{1-\epsilon}\over{V_n}}.
\label{c10_eq13}
\end{eqnarray}

\begin{figure}[t]
\centerline{\includegraphics[width=1\linewidth]{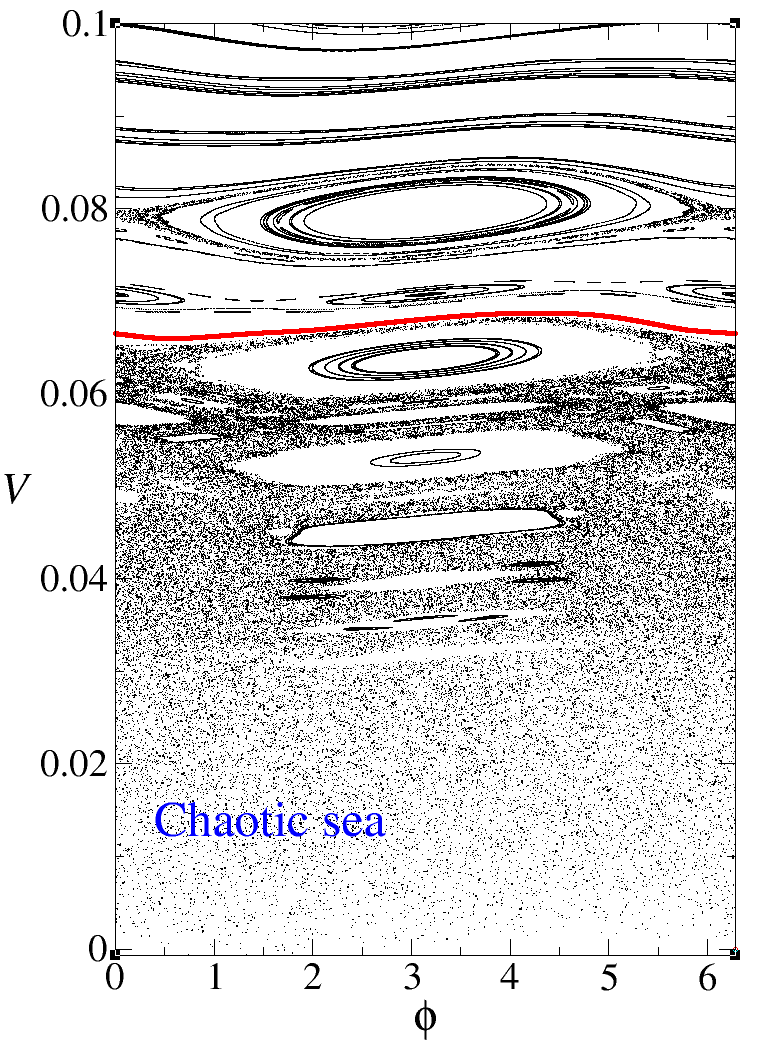}}
\caption{Plot of the phase space for the mapping (\ref{fum_eq}) using $\epsilon=10^{-3}$. The mixed structure is observed where chaos surrounds periodic islands and is limited by a first invariant spanning curve.}
\label{Fig1}
\end{figure}

The two functions $F$ and $G$ are solved using a very accurate numerical method \cite{hamming2012numerical} called the bisection method, where the solution converges to a tolerance of $10^{-12}$ and there is no excessive computational time. 
A plot of the phase space for the parameter $\epsilon=10^{-3}$ is shown in Figure \ref{Fig1}. 

It is evident the existence of the periodic islands, which are centered by elliptic fixed points embedded in a chaotic sea, is determined (see Ref. \cite{leonel2021scaling}) by a positive Lyapunov exponent of $\lambda=0.728(1)$. We use a triangularization algorithm that transforms a square $2~\times~2$ non-null elements Jacobian matrix into a new one in terms of a triangular shape \cite{eckmann1985ergodic}. A first invariant spanning curve limits the chaotic sea. It has a similar behavior of a barrier and does not allow the crossing of any particle through it, hence defining a critical scaling for the chaotic sea \cite{leonel2004fermi}. It is proven that a more stable invariant spanning curve \cite{joelson} has returning times whose number belongs to the Fibonacci sequence. It is the last one to be destroyed as the control parameter is varied.

In the next section, we discuss the occurrence of multiple collisions. They are observed in the chaotic sea for the regime of low energy, hence for the domain where the particle's velocity is comparable with the maximum velocity of the moving wall. 

\begin{figure}[t]
\centerline{\includegraphics[width=1\linewidth]{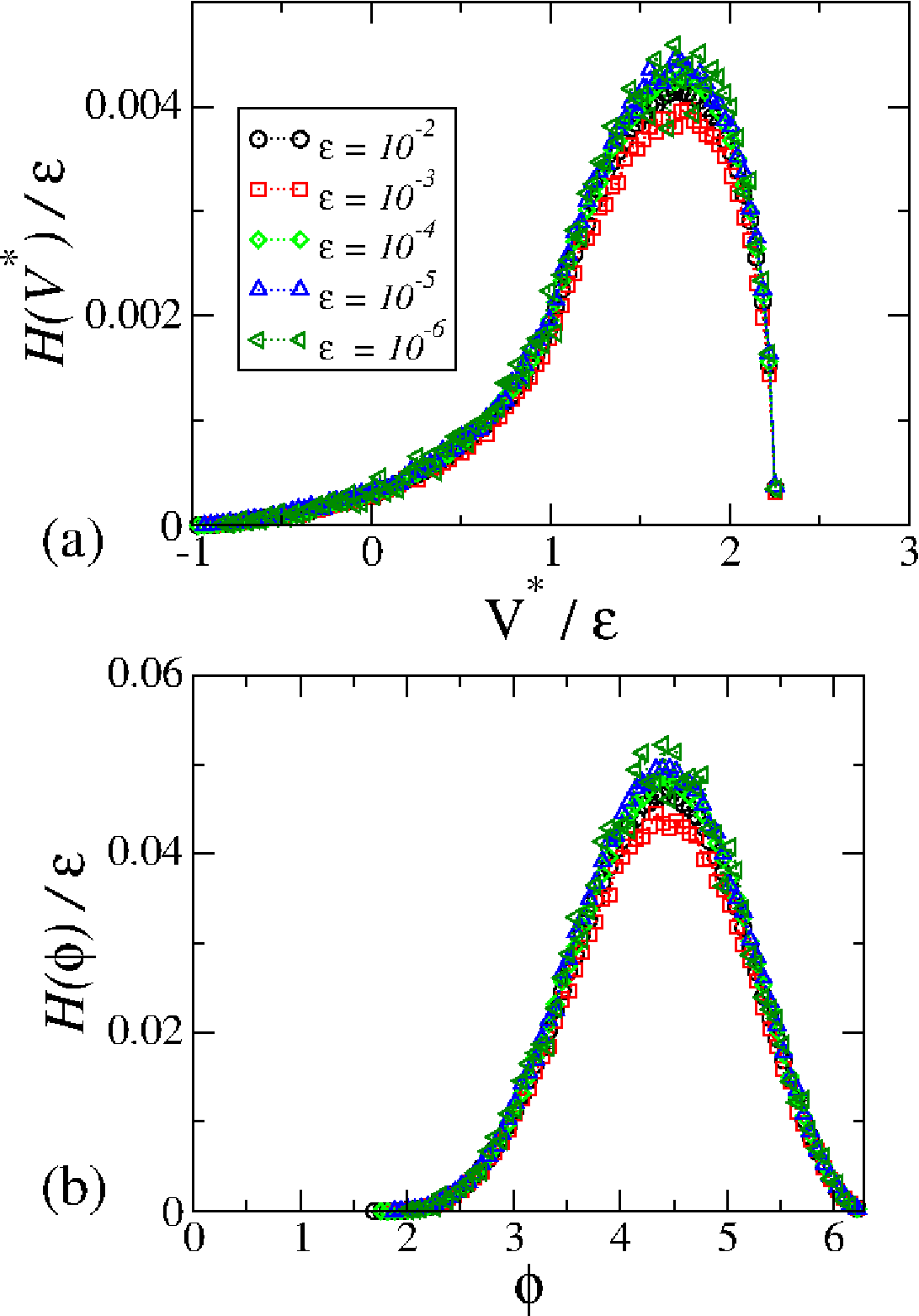}}
\caption{Plot of $p~vs.~\epsilon$ obtained from Eq. (\ref{p}).}
\label{fig2}
\end{figure}

\section{Distribution of the multiple collisions}
\label{sec3}

For the regime of low energy and depending on the velocity of the particle and the phase of the moving wall, the particle may enter the collision zone and, after experiencing a collision with the moving wall and before it leaves the collision zone, it may experince many other collisions. Since the particle is moving in the chaotic domain of the low energy regime, we may assume that the particle has a probability $p$ to move to the right after the collision while $q$ is the probability it moves to the left. The probability of an event is then the product of the two and is written as $P=pq$. However, the particle may have yet the probability of experiencing $n_1$ other collisions moving to the right and $n_2$ collisions moving to the left. This is possible because the position of the moving wall is accelerated. If the particles experince a total of collisions $N$, $n_1$ of them can lead the dynamics to the right while $n_2$ leads to a motion to the left, and the total probability is $P_N=p^{n_1}q^{n_2}$. We may note that many possible distinct combinations of $ n_1 $ and $ n_2 $ lead to a total of $ N $ successive impacts. The multiple possibilities of the sequence are taken into account with the following combinatory, yielding a binomial distribution for the probability
\begin{equation}
P_N={{N!}\over{n_1!(N-n_1)!}}p^{n_1}q^{N-n_1},
\label{comb}
\end{equation}
where $n_2=N-n_1$.

\begin{figure}[t]
\centerline{\includegraphics[width=1\linewidth]{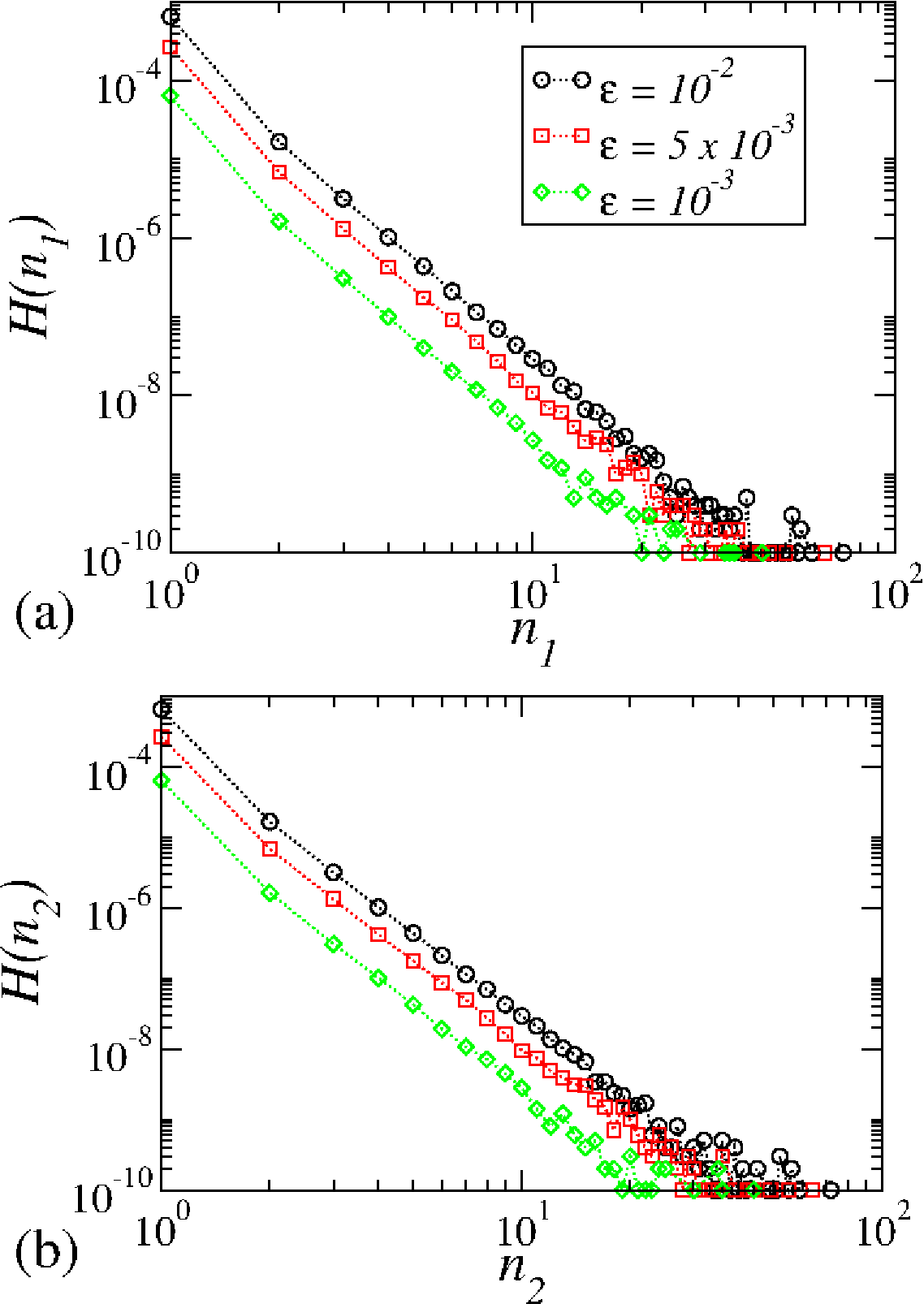}}
\caption{Plot of the probability distribution for the successive reflection for (a) $n_1$ and (b) $n_2$. The control parameters are shown in the figure.}
\label{Fig3}
\end{figure}

\begin{figure}[t]
\centerline{\includegraphics[width=1\linewidth]{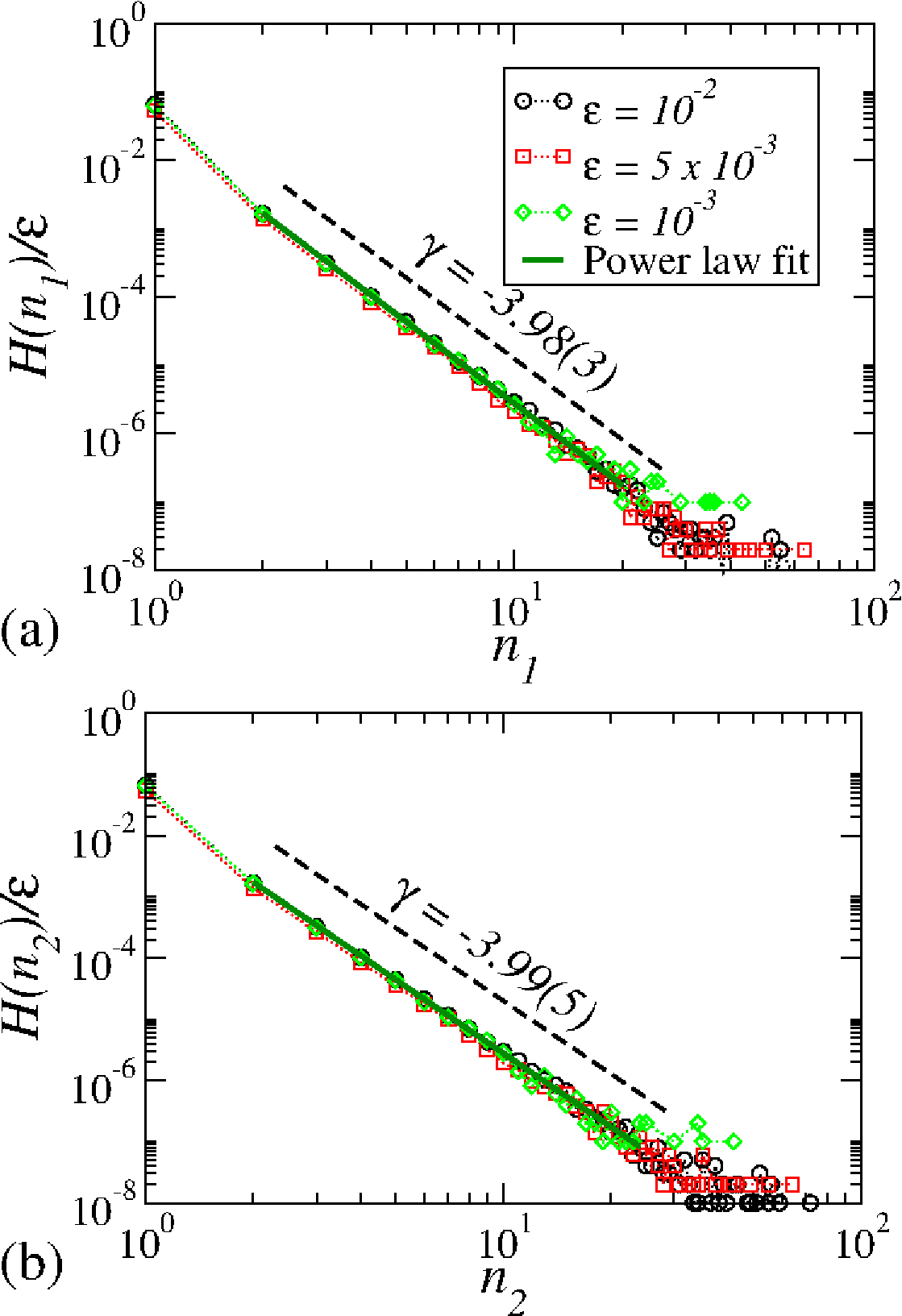}}
\caption{Plot of the normalized probability distribution for the successive reflection for (a) $n_1$ and (b) $n_2$. The control parameters are shown in the figure.}
\label{Fig4}
\end{figure}

When the particle enters the collision zone with $V^*\le\epsilon$, a probability that it collides can be estimated by the ratio of the difference of velocity by two times the amplitude of the moving wall velocity, i.e.
\begin{equation}
p=q={{\epsilon-V^*}\over{2\epsilon}},
\label{p}
\end{equation}
where $V^*\in[-\epsilon,\epsilon]$. Figure \ref{fig2}(a) shows a plot of the distribution for the velocity in the successive collisions
while Fig. \ref{fig2}(b) shows the histogram with the most frequent phases in which successive collisions happen. The axis of Fig. \ref{fig2}(a) were chosen such $V^*\rightarrow V^*/\epsilon$ and $H\rightarrow H/\epsilon$, and we see all curves are fitted together into a single plot that depends on the control parameter. The shape of the curve shown in Fig. \ref{fig2}(a) is remarkably similar to a binomial distribution, as foreseen by Eq. (\ref{comb}). The histogram showing the window of time where the successive collisions happen, as shown in Fig. \ref{fig2}(b), is rescaled in the vertical axis to show independence of the control parameters for the following transformation $H\rightarrow H/\epsilon$, overlapping all curves obtained for different control parameters onto a single plot.

The main question to be answered is, after the particle enters the collision zone and experience $N$ successive impacts, what are the corresponding average values for $\overline{n}_1$ and $\overline{n}_2$? The particle may then leave the collision zone, move around the chaotic sea, and come again for a further set of successive collisions with the same probability. We can define the number of times the particle entered the collision zoned with the probability of having $N=1,2,3,\ldots$ successive reflections. The cumulative distribution function for the successive collisions can be obtained by
\begin{equation}
H(n_1)=\sum_{n_1=1}^NP_N(n_1).
\label{cumulative}
\end{equation}

From the mapping (\ref{fum_eq}), we can start an initial condition and follow a very long orbit in the chaotic attractor, typically of $10^{11}$ collisions. The particle will move around the chaotic sea, and from time to time, it will experience multiple collisions. The numerical distribution for the successive collisions is shown in Fig. \ref{Fig3}(a) for $n_1$ and (b) for $n_2$, considering different control parameters.

We notice the behavior of the probability distribution function decays rapidly in a log-log plot showing a power law behavior as $F(n)\propto n^{\gamma}$. A power law fitting gives an exponent similar to the two distributions. The exponents obtained for the decay in $n_1$ is $\gamma=-3.98(3)$ while for $n_2$ is $\gamma=-3.99(5)$.

We also notice the control parameter affects the position where the curves start to decay. A straightforward transformation of $H(n)\rightarrow H(n)/\epsilon$ overlaps all curves onto a single and universal plot as shown in Fig. \ref{Fig4} for the same set of control parameters used for Fig. \ref{Fig3}.

The chaotic dynamics for particles moving along the phase space are already known to be scaling invariant \cite{leonel2004fermi}. The Lyapunov exponents calculated led to a convergence to a stationary state \cite{leonel2004dynamical}, proving also to have almost no dependence on the control parameters. The existence of the first invariant spanning curve has a fundamental role in limiting the size of the chaotic sea. It can be determined as a function of the control parameter $\epsilon$, yielding the existence of a set of critical exponents. When they are used to define scaling variables, all curves of the average velocity, which gives the diffusion in the phase space, are overlapped onto a single and universal plot, which is proved to be the scaling invariant for the control parameters \cite{leonel2004fermi}. The present result also gives clear evidence that the chaotic dynamics for the very low energy regime for the bouncing ball model is also scaling invariant. The number of successive collisions of the particle with the moving wall is clearly described by a power law decay with an exponent of $\gamma=-4$.

\section{Conclusion}
\label{conc}

In summary, we have revisited the bouncing ball model to describe the behavior of a rare set of collisions presented in the low chaotic energy domain called successive collisions. They happen when the particle enters the collision zone with low energy. Since the motion of the moving wall is accelerated, the particle may experience successive collisions inside the moving wall before leaving from there. Our results show the probability of observed multiple collisions when the particle moves to the right or left, described by a power law with an exponent $\gamma-4$. The histogram shows the frequency of successive velocities.

\section*{Acknowledgements}
E.D.L. acknowledges support from Brazilian agencies CNPq (No. 301318/2019-0, 304398/2023-3) and FAPESP (No. 2019/14038-6 and No. 2021/09519-5)


\end{document}